\documentclass[a4paper,twocolumn,english,aps,prl,reprint,superscriptaddress,showpacs,showkeys,longbibliography]{revtex4-1}
\usepackage[latin9]{inputenc}
\usepackage{babel}
\usepackage{amsmath}
\usepackage{amssymb}
\usepackage{graphicx}
\usepackage[colorlinks=true, linkcolor=red, citecolor=blue, urlcolor=blue]{hyperref}
\usepackage[usenames,dvipsnames]{color}
\usepackage{braket}
\usepackage{float}
\usepackage{comment}

\definecolor{mygreen}{rgb}{0,0.5,0}
\definecolor{myblue}{rgb}{0,0,0.75}
\definecolor{mymagenta}{cmyk}{0,1,0,0.12}

\newcommand{\minus}{
  \setbox0=\hbox{-}
  \vcenter{
    \hrule width\wd0 height \the\fontdimen8\textfont3
  }%
}

\begin{document}

\title{Quantum Approximate Optimization with Parallelizable Gates}

\author{Wolfgang Lechner}
\affiliation{Institute for Theoretical Physics, University of Innsbruck, A-6020
Innsbruck, Austria}
\affiliation{Institute for Quantum Optics and Quantum Information of the Austrian
Academy of Sciences, A-6020 Innsbruck, Austria}

\date{\today}

\begin{abstract}
The quantum approximate optimization algorithm (QAOA) has been introduced as a heuristic digital quantum computing scheme to find approximate solutions of combinatorial problems with shallow circuits. We present a scheme to parallelize this approach for arbitrary all-to-all connected problem graphs in a layout of quantum bits (qubits) with nearest neighbor interactions. The protocol consisting of single qubit operations that encode the optimization problem and all interactions are problem-independent pair-wise CNOT gates among nearest neighbors. This allows for a parallelizable implementation in quantum devices with a square lattice geometry. The basis of this proposal is a lattice gauge model which also introduces additional parameters and protocols for QAOA to improve the efficiency. 
\end{abstract}

\pacs{}

\keywords{}

\maketitle

\textit{Introduction} - With the immense recent developments in quantum technology \cite{georgescu2014quantum,wallraff2004strong,koch2007charge,cirac2012goals,blatt2012quantum,bloch2012quantum,houck2012chip,bravyi2017quantum,otterbach2017unsupervised,bernien2017probing,watson2017programmable,zhang2017observation} the regime of computational quantum advantage is in reach \cite{preskill2018quantum,boixo2016characterizing,harrow2017quantum}. Solving computationally hard optimization problems is a possible application of such near-term intermediate-size special purpose quantum computers that currently receives  considerable interest. The working principle to solve optimization problems on quantum hardware is to encode the problem in a spin model such that the cost function corresponds to the energy of the system \cite{lucas2014}. Finding the ground state of the spin model is thus equivalent to solving the optimization problem. In general, hard problems translate to disordered all-to-all connected Ising spin glasses \cite{kirkpatrick1983optimization}. Due to the large number of local minima it is computationally challenging to find the ground state of such a model using current state of the art classical algorithms \cite{ingber1993simulated,KatzgraberSA,heim2015quantum}. 

The quantum approximate optimization algorithm (QAOA) \cite{farhi2014quantum,farhi2016quantum} has been recenlty introduced as a heuristic digital quantum algorithm to sample approximate ground states using shallow quantum circuits. The scheme consists of a sequence of quantum quenches represented by unitary operators that correspond to a driver Hamiltonian and a problem Hamiltonian, respectively. The number of iterations can be small and angles of each unitary are free parameters that are optimized via a classical feedback loop. The method has been recently proven to exhibit the optimal Groover speedup in unstructured search \cite{QAOARieffel}. An open challenge is scalability and programability to encode arbitrary optimization problems independent of the physical qubit arrangement and connectivity. While larger connectivity can be achieved by a series of swap operations, these operations are problem-dependent and thus difficult to parallelize which is a limiting factor in scalability and execution speed. 

\begin{figure*}[htb]
	\centering
\includegraphics[width= 15cm]{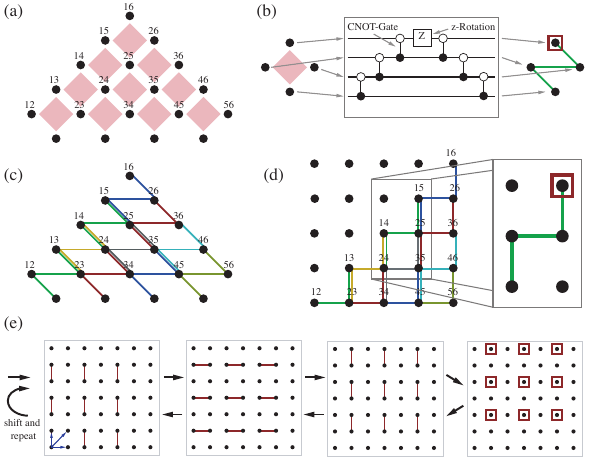}
\caption{(a) Lattice gauge formulation of an all-to-all connected spin model. The z-component of  the qubits (dots) represent relative coordinations of two spins $i$ and $j$ (labels) and the optimization problem is encoded in the local fields acting on individual qubits. The interactions are problem-independent 4-body interactions on a square lattice (red). (b) The QAOA unitary to implement a LHZ plaquette interaction is decomposed into a series of 3 CNOT gates along a z-shaped path followed by a qubit rotation and 3 CNOT gates in reversed order back. (c) Pattern of CNOT gates to implement all 4-body interactions. (d) The system can be realigned to a regular square lattice. In the square lattice, the LHZ 4-body plaquette interactions decompose into a sequence of gates that connects two physical plaquettes with CNOT gates (inset, green) and a single $R_Z$-Gate (inset, square) that determines the strength of the constraint. (e) Sequence of parallel CNOT gates (lines) and $R_Z$ rotations (squares) to realize all 4-body interactions consisting of $7$ parallel gate operations. The pattern is then shifted up by one row, shifted to the right by one column and shifted up-and-right (blue arrows) and repeated. Thus a total number of 28 parallel gates is required to realize all constraints independent of the system size.}
\label{fig1}
\end{figure*}

In this work, we present a parallelizable QAOA scheme consisting of nearest-neighbor CNOT gates and single qubit rotations with the aim to solve all-to-all connected combinatorial optimization problems. The scheme is based on the recently introduced encoding of optimization problems in a lattice gauge model (LHZ) \cite{lechner2015}. In this mapping, the problem is fully determined by local fields while interactions are uniform and problem-independent. This separation applied to QAOA allows for an implementation with pair-wise gates that are executed in parallel on a square lattice with nearest-neighbor connectivity. The required gates consists of three terms: (i) a unitary with local $\sigma_x$ terms, (ii) a unitary that defines the problem with local $\sigma_z$ terms, and (iii) problem-independent interactions consisting of nearest-neighbor CNOT gates and qubit rotations, illustrated in Fig. \ref{fig1}. The mapping also introduces additional free parameters and new QAOA protocols that may be used to increase the efficiency of the method.

\textit{Lattice Gauge QAOA} - The quantum approximate optimization algorithm \cite{farhi2014quantum} aims at finding approximate solutions in a hybrid quantum-classical approach inspired by quantum annealing  \cite{Nishimori1998,Farhi2000,albash2016adiabatic,Boixo2014,Boixo2016}. In quantum annealing, the ground state of the problem Hamiltonian is prepared by an adiabatic sweep of the form $H(t) = \frac{t_{\textrm{max}}-t}{t_{\textrm{max}}} H_0 + \frac{t}{t_\textrm{max}} H_p$, where $H_0 = \sum_i \sigma_x^{(i)}$ is the driver Hamiltonian, and $H_p = \sum_{i<j} J_{ij} \sigma_z^{(i)}\sigma_z^{(j)}$ is the problem Hamiltonian. The system is initially prepared in the ground state of $H_0$. Given that $t_{\textrm{max}}$ is large compared to the minimal gap in the time-dependent spectrum the system will remain in the instantaneous ground state and thus the system is found in the ground state of $H_p$ at time $t_{\textrm{max}}$. It is an open question whether a quantum speedup for hard combinatorial problems can be expected with this protocol \cite{albash2016adiabatic,Boixo2014}.

In QAOA, instead of adiabatically transforming the Hamiltonian, the system is sequentially quenched e.g. with 
\begin{equation}
	\label{eq:psi}
	|\psi(m,\beta_1, \gamma_1,...) \rangle = U_x(\beta_1)U_p(\gamma_1) ... U_x(\beta_m)U_p(\gamma_m) |s\rangle.  
\end{equation}
Here, $| s \rangle $ is the initial state which is typically chosen to be the equal superposition in computational basis $| s \rangle = \frac{1}{\sqrt{2^N}} \sum | z \rangle $  and the unitary operators
\begin{equation}
	\label{eq:ud}
	U_x(\beta) = e^{- i \beta H_0} = \prod_{i=1}^N e^{-i\beta \sigma_x^{(i)}},
\end{equation}
and 
\begin{equation}
	\label{eq:ud}
	U_p(\gamma) = e^{- i \gamma H_p}. 
\end{equation}
The number of iteration cycles is $m$ as well as $\beta_i$ and $\gamma_i$ are free parameters that are tuned to minimize the expectation of the final state with respect to the problem Hamiltonian
\begin{equation}
E = \textrm{min}_{\gamma,\beta} \langle \psi | H_p | \psi \rangle.
\end{equation}
This parameter optimization is done via a hybrid quantum-classical algorithm where the state $|\psi \rangle$ is prepared using the quantum device and the parameters are updated as classical feedback from the outcome of measurements. For optimization problem, one might be interested in the probability to find the best solution which is measured by the ground-state fidelity
\begin{equation}
F =  \langle \psi | \varphi_0 \rangle.
\end{equation}
Here, $|\varphi_0 \rangle$ is the ground-state of $H_p$. The challenge that we address here is the implementation of $U_p$ which consists of geometrically non-local terms that can, in general, not be  mapped directly to the physical qubit layout. 

The non-local and disordered interactions can be removed with the recently introduced LHZ mapping \cite{lechner2015}, which is a lattice gauge formulation for optimization problems. The LHZ-Hamiltonian has the form \cite{rocchetto2016stabilizers,albash2016simulated,pastawski2016error,puri2017quantum,leib2016transmon,puri2017quantum,glaetzle2017coherent}
\begin{eqnarray}	
	\label{eq:lhzh}
	H(t) &=& H_0(t) + H_p(t) \textrm{, where} \\
	H_0(t) &=& A(t) \sum_i^K \sigma_x^{(i)} \\ 
	H_p(t) &=& B(t) \sum_i^K J_i \sigma_z^{(i)} + \\ \nonumber
	&+& C(t) \sum_{l=1}^{K-N+1} C_l \sigma_z^{(l,n)}\sigma_z^{(l,e)}\sigma_z^{(l,s)}\sigma_z^{(l,w)}
\end{eqnarray}
Here, $K= N(N-1)/2$ is the number of connections in the graph and the indices $n,e,s,w$ denote north, east, south and west qubit of each plaquette. The schedule function $A$ is switched from 1 to 0 and $B$ and $C$ are switched from 0 to 1, respectively. The physical qubits represent the relative orientation of the spins in the optimization problem. Thus the number of physical qubits is the number of edges in the problem graph, which introduces a quadratic overhead for general all-to-all models. The matrix elements $J_{i}$ that encode the optimization problem are, in contrast to the spin model, local fields acting on single qubits. A third, additional term is introduced that energetically constrains the system in the increased Hilbert space to the low energy sub-space. The constraints are 4-body interactions acting on plaquettes [see Fig. \ref{fig1}(a)] of a square lattice. 

Using the above mapping in the spirit of QAOA suggests the following unitary operators as building blocks for the optimization algorithm:
\begin{eqnarray} \label{eq:UX}
U_x(\beta) &=& \prod_{i=1}^K e^{-i\beta \sigma_x^{(i)}}, \\ 
\label{eq:UP}
U_z(\gamma) &=& \prod_{i=1}^K e^{-i\gamma J_i \sigma_z^{(i)}}, \textrm{and} \\
\label{eq:UC}
U_c(\Omega) &=& \prod_{l=1}^{K-N+1} e^{-i\Omega C_l \sigma_z^{(l,n)}\sigma_z^{(l,e)}\sigma_z^{(l,s)}\sigma_z^{(l,w)}}.
\end{eqnarray}
 The unitary $U_x$ is the propagator of the driver Hamiltonian. The problem Hamiltonian is now split into two parts, $U_z$ and $U_c$. In the latter, $C_l$ are the constraint strengths of each plaquette $l$. The constraint strengths are free parameters because the low energy subspace is gauge invariant under change in $C_l$. They can be optimized in addition to the angles $\beta$, $\gamma$, $\Omega$. With the separation of interactions and local fields, the terms in $U_z$ and $U_x$ are all simple single qubit rotations and phase rotations. The only programmable (and therefore disordered) Hamiltonian is $U_z$. The remaining term containing the interactions (i.e. $U_c$) is problem-independent. Due to this independence of interactions and encoded problem the two-qubit gates are uniform and in the following a parallelizable implementation is discussed. 
 
 The 4-body interactions required in $U_c$ are problem independent and thus identical for each plaquette. These individual plaquette terms be can realized as shown in Fig. \ref{fig1}(b) using 6 CNOT gates and one qubit rotation $R_z$. The CNOT gates are performed along a \textit{path} connecting all 4 qubits, followed by a qubit rotation is performed followed by the same CNOT gates performed in reverse order. We have chosen the particular z-shaped path as shown in Fig. \ref{fig1}(b). Note, that other \textit{paths}, e.g. clockwise along a plaquette is also possible. Plugging this gate sequence into the LHZ architecture results in the particular connectivity graph as shown in Fig. \ref{fig1}(c). By realignment to a regular square graph, the CNOT gates act on nearest neighbors only [see Fig. \ref{fig1}(d)]. Note, that the plaquette interactions in LHZ translate to a sequence of CNOT gates along lines that connect two plaquettes in the physical graph [see Fig. \ref{fig1}(d) (right)]. As these interaction are identical for each plaquette, they can be executed in parallel. Fig. \ref{fig1}(e) shows a sequence with parallel gates consisting of 3 particular CNOT gates, a $R_z$ gate and the 3 CNOT gates in reversed order. To reach all plaquettes, this sequence is repeated in total 4 times, where after each run the gates are shifted by one row, one column and one row-and-column. This makes in total 28 parallel gates to implement all constraints. 

Note, that the strength of the constraints $C_l$ are determined by the $R_Z$-Gate alone and that the CNOT gates are independent of the problem and also independent of the constraints. Thus, only local Z operations contain disorder and all CNOT gates are problem-independent and parallelizable on a two dimensional grid. 

\textit{Parameter and Protocol Optimization} - QAOA is a feedback driven algorithm \cite{farhi2014quantum}. The feedback consists of measuring the outcome of quenches according to a protocol in Eq. \eqref{eq:psi} on the quantum device and using classical optimization methods to improve the parameters $\beta_i$, $\gamma_i$ in Eq. \eqref{eq:ud}. In the protocol presented here, the unitary operators given in Eqs. \eqref{eq:UX}-\eqref{eq:UC} introduce additional free parameters ($\Omega_i$ and $C_l$) and additional quench protocols. In the following we address the question, how these additional degrees of freedom can be used to improve the algorithm.

The unitary operators Eqs. \eqref{eq:UX}-\eqref{eq:UC} allow for two possible algorithmic directions of improvement: i.) the protocol that determines the order and form of the operators and ii.) the  choice of parameters that are varied. In order to compare the different approaches we keep the number of feedback iterations $m$ fixed as this is the limiting factor in experiment. The initial state is chosen to be the uniform superposition in the computational basis and the total quench is denoted as $| \psi \rangle = U_{a,b,c} |s \rangle$. We consider for illustration 3 particular protocols: 
\begin{eqnarray}
	&U_a& = U_{p}(\gamma_0) U_{x}(\beta_1)U_{p}(\gamma_1) ... \\
	&U_b& = U_{z}(\gamma_0) U_{c}(\Omega_0) U_{x}(\beta_1) U_{z}(\gamma_1) U_{c}(\Omega_1) ... \\	
	&U_c& = U_{z}(\gamma_0) U_{c}(\Omega_0,C_l) U_{x}(\beta_1) U_{z}(\gamma_1) U_{c}(\Omega_1,C_l) ... 	
\end{eqnarray}
Here, protocol $U_a$ is a sequence of applying the LHZ problem Hamiltonian with $U_{p} = e^{-i\gamma H_p}$ and the driver Hamiltonian. Protocol $U_b$ makes use of the splitting between local field terms and interaction terms and optimizes the parameters $\beta$,$\Omega$ and $\gamma$ independently. Protocol $U_c$ also includes an update of the constraint strengths and thus the Hamiltonian itself.

We consider an optimization problem encoded in $K = 6$ qubits arranged on a square lattice with $3$ plaquettes. The parameters are optimized using the following Monte Carlo procedure:  1.) The parameters are initialized with $\gamma_i,\beta_i,\Omega_i = 1$ and $C_l=2$. The interaction matrix is chosen randomly from the interval $J_{ij} \in \{-1,...,1\}$.  2.) The final states $|\psi\rangle$ are prepared according to the above protocols $U_a$,$U_b$ and $U_c$. 3.) The expectation $E$ and the fidelity $F$ are determined. 4.) This is repeated $M=4000$ times and in each Monte Carlo step a randomly chosen parameter is updated by a random number in the range $\delta = \{-1,...,1\}$. The set of parameters is accepted if the expectation $E$ decreases and rejected otherwise. For comparison we also optimized parameters directly improving the fidelity $F$. In this case, the update is accepted if the fidelity increases. 5.) This procedure is repeated for random instances of $J_{ij}$. The averages are taken over $L=2000$ realizations. The 3 protocols $U_a$, $U_b$ and $U_c$ are compared using the same instances for numbers of iteration $m=1,2$ and $3$. In protocol $U_c$ the update of $C_l$ is attempted every 10th steps. 

\begin{figure}[htb]
	\centering
\includegraphics[width= 8.5cm]{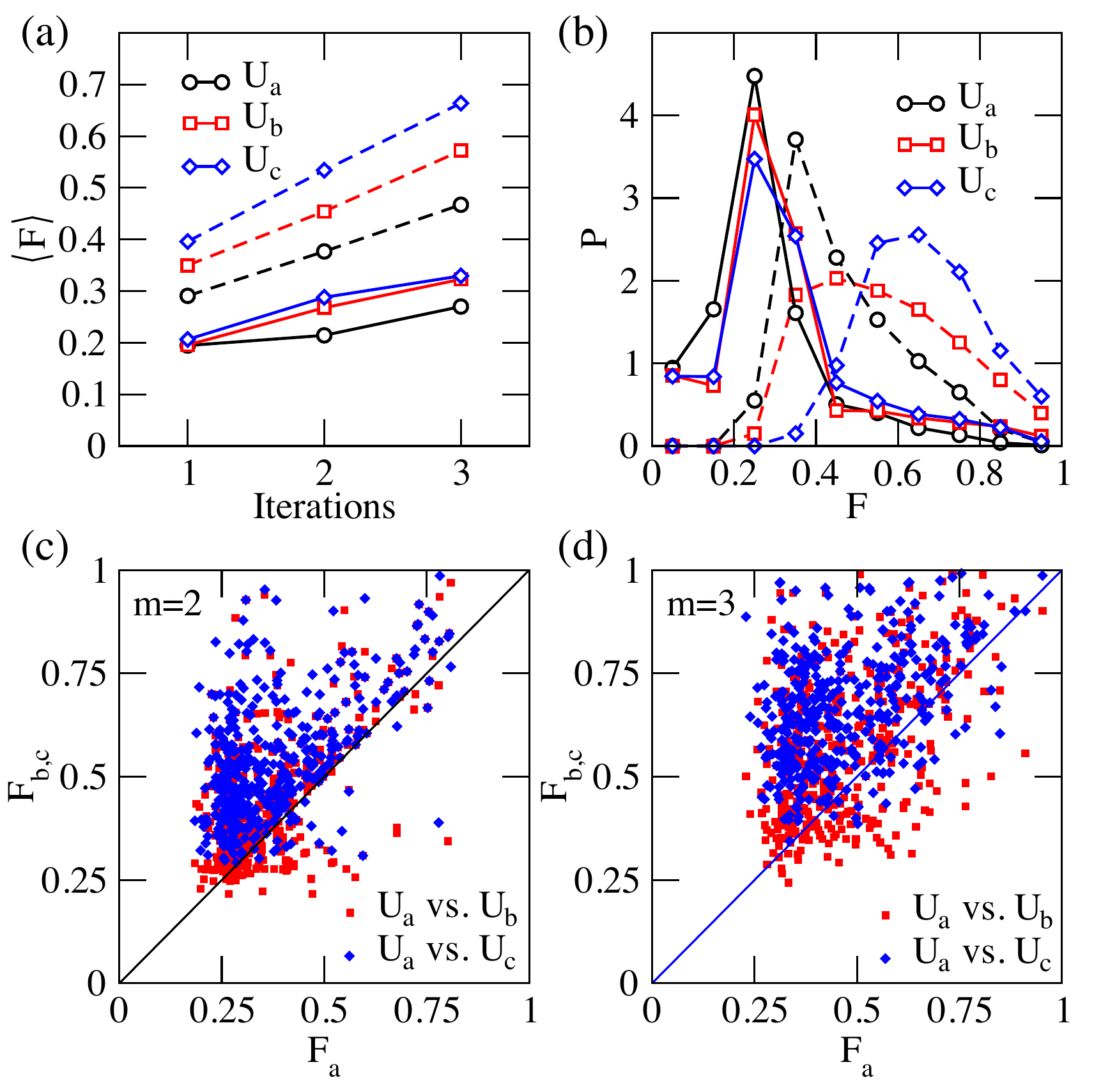}
\caption{(a) Average fidelity defined as overlap of the final state with the ground-state of the problem Hamiltonian as a function of number of iteration cycles. Average is taken over $L=400$ random instances optimizing $F$ (dashed) and $L=2000$ instances optimizing the expectation $E$ (solid). QAOA with driver and problem Hamiltonian (protocol $U_a$, black) improves with number of iterations. Separation of the problem Hamiltonian (protocol $U_b$, red) and optimizing also the constraints (protocol $U_c$, blue) improves the fidelity for all $m$, the number of iterations. (b) Normalized histogram $P$ of the fidelity for the same parameters as in panel $m=3$ iterations optimizing the expectation (solid) and the fidelity (dashed). (c) Scatter plot of the fidelity optimization with same parameters as in panel a with $m=2$ iterations comparing direct fidelity optimization with protocol $U_a$ against protocol $U_b$ (red) and against protocol $U_c$ (blue), respectively. (d) Scatter plot of the fidelity comparing protocols as in panel c with $m=3$ iteration cycles. }
\label{fig2}
\end{figure}

Figure \ref{fig2} depicts the fidelity in comparison for protocols $U_a$, $U_b$, and $U_c$ [see Fig. \ref{fig2}(a)]. The separation of local field terms and interaction terms is advantageous, and the additional optimization of constraints using protocol $U_c$ does further improve the fidelity. Note that this protocol may result in a further improvement if more measurements are used. The histograms of the fidelities for the various protocols show that protocol $U_c$ has the best average performance and also the largest contributions for fidelities close to $F=1$ [see Fig. \ref{fig2}(b)]. A direct comparison of protocols [see Fig. \ref{fig2}(c) and (d)] show that most instances result in a lower fidelity using protocol $U_a$ compared with the optimal protocol $U_c$. 

\textit{Conclusions and Outlook - } We have applied the LHZ mapping to QAOA which allows one to separate the two-qubit gates from the programmable local fields containing the optimization problem. With this separation we introduce a scheme with full  parallelization of gates on the quantum device. The optimization problem is solved with problem-independent CNOT gates that can be performed in parallel on a square lattice while only local fields are problem dependent, which is directly applicable to qubits on a two-dimensional grid \cite{boixo2016characterizing}. The mapping also suggests several novel QAOA protocols, in particular optimization of local fields independent of the optimization of interactions. Using Monte Carlo as an optimization technique and comparing 3 particular protocols with fixed number of readouts, we find that for the given instances the best protocol is to use the unitary operators as given in Eqs. \eqref{eq:UX}-\eqref{eq:UC} and an update of the angles and gauge constraints. Note, that the depicted geometry in Fig. \ref{fig1} corresponds to an all-to-all graph. The layout represents a restricted Boltzmann machine if the full square lattice is filled, a mapping that could be directly applicable for unsupervised machine learning applications \cite{otterbach2017unsupervised}.

As a future direction, the parallelizable QAOA scheme may be extended to more complex driver Hamiltonians e.g. non-stoquastic terms \cite{bravyi2006complexity,nishimori2017exponential,hormozi2017nonstoquastic}.

\textit{Acknowledgements.---} The author thanks E. Farhi for fruitful discussions. Research was funded by the Austrian Science Fund (FWF) through a START grant under Project No. Y1067-N27, the Hauser-Raspe foundation. The computational results presented have been achieved using the HPC infrastructure LEO of the University of Innsbruck.

\end{document}